# Temperature of a spinning black hole via a simple derivation

Ronald J. Adler

Gravity Probe B mission, retired
Stanford University,
Stanford CA 94035

Department of Physics and Astronomy, retired
San Francisco State University,
San Francisco CA 94132

gyroron@gmail.com

**Abstract**

According to current theory a black hole has a nonzero temperature and thus radiates like any black body. This remarkable result was first shown by Hawking for a non-spinning black hole using general relativity to describe the black hole gravitational field and quantum field theory to describe the radiation. Since then the temperature of a spinning Kerr black hole has been calculated. There have also been many heuristic derivations for the temperature. In this work we derive the temperature of a Kerr spinning black hole using only classical general relativity and thermodynamics. It is very similar to Ref. 11 but is mathematically simpler and more self-contained. Our purpose is mainly pedagogical, to be more accessible to students and non-specialists with a knowledge of general relativity. We also call further attention to the expected explosive evaporation of small black holes, not yet observed, which would be an almost unique window into Planck scale physics. Finally, we discuss the idea that the cosmological dark matter, whose nature is currently unknown, may be composed of small primordial black hole remnants.

# 1. Introduction

Hawking was the first to obtain the extraordinary theoretical prediction that non-spinning black holes (BHs) have a nonzero temperature and thus radiate like a black body. His calculation used classical general relativity (GR) and quantum field theory.[1,2] Since then the result has been generalized to spinning BHs, calculated in a similar manner.[3, 4, 5] There have also been numerous heuristic derivations that serve to clarify the basic physics behind the process.[6,7,8] It is worth noting that a purely dimensional argument does *not* work.

Our derivation of the temperature of a spinning BH depends only on classical GR and on black body thermodynamics.[6,9] These two inputs are among the best understood and tested in all of physics. In particular this derivation should be accessible to those with an understanding of GR. The price we pay is that our result is approximate, with an overall multiplicative constant of order 1.

The derivation is based on that of Ohanian and Ruffini, but generalizes their work to spinning BHs[10,11]. It is very similar to Ref. 11 but deals with the polar axis region of the spinning BH geometry rather the equatorial plane.[11] It thereby makes the calculation simpler and self-contained, as well as showing that the resultant temperature describes the entire BH rather than separate regions. We also discuss more completely the potential function used in the energy calculations.[11]

In section 2 we set the stage for the GR calculation by doing a simple thought experiment in the context of Newtonian theory: Fig. 1 shows how we can do work and extract energy by lowering a small body to near a massive spherical body.

In section 3 we do the same experiment using GR, as shown in Fig. 2. It is important that we lower the small body along the north polar axis.[6] The interesting quantity is the energy that can be extracted for distances very close to the BH null surface. This section is self-contained and accessible to those familiar with GR and the Kerr metric.

In section 4 we view the apparatus in Fig. 2 as a Carnot heat engine, using a hot thermal reservoir in the distant lab and the BH as a cold reservoir, as shown in Fig. 3. The Carnot theorem of thermodynamics for the efficiency, combined with the result of section 3, then gives the BH temperature.[9] The result is approximate in that it refers to a "characteristic wavelength" for the radiation in the black body.[11] Our final result for the BH temperature is summarized in (4.7) and shown in Fig. 3.

In section 5 we discuss why the temperature of a BH is relevant to fundamental physics: very low mass and thus very hot BHs should evaporate explosively while emitting all the particles in nature – giving an almost unique window on Planck scale physics! [12]

Moreover, if there are remnants from such explosions they would be a possible candidate for the cosmological dark matter. Alas no such explosions have yet been observed.[13]

# 2. Energy extraction with a spherical Newtonian body

In the next two sections we will calculate how much energy we can extract from gravity by lowering a small body onto or nearly onto a massive body, in particular a BH. We first set the stage with the simple case of a massive sphere in Newtonian theory, and in the next section we will repeat the calculation for a BH using GR. Fig. 1 shows our thought experiment, in which a small box of mass $\mu$ is lowered slowly, with a rope, from a large distance where the potential energy $V$ is negligible, toward the body of mass $M$; the geometric mass is defined as $m = GM/c^2$ and the radius is taken to be $r_s = 2m$.

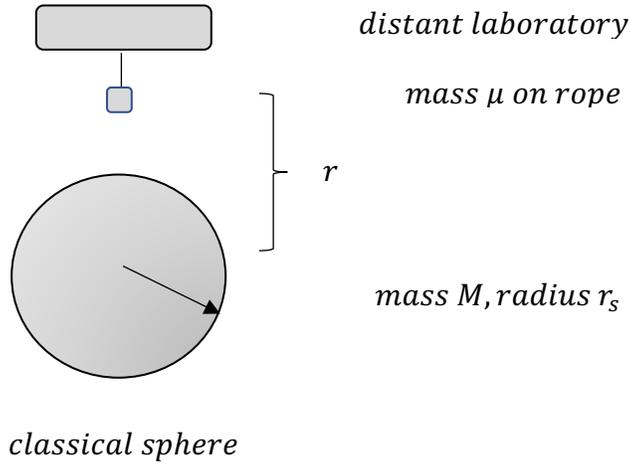

Figure 1. Thought experiment to extract energy from gravity, using Newtonian theory.

The rope pulls downward and does work $W$ that is the difference in Newtonian potential energy $V$,

$$V(r) = -\frac{GM\mu}{r} = -\frac{m}{r}\mu c^2, \quad m = \frac{GM}{c^2}, \tag{2.1}$$

$$W = V(\infty) - V(r) = \frac{GM\mu}{r} = \frac{m}{r}\mu c^2. \quad \text{(Newtonian)} \tag{2.2}$$

If we lower the box to the surface at $r_s = 2m$ we thus obtain work

$$W = \frac{\mu c^2}{2}, \quad \text{(half the rest energy at the surface).} \tag{2.3}$$

If we only lower the box to near the surface, $r = 2m + \Delta l$, we have from (2.2)

$$W = \frac{\mu c^2}{2}\left(1 - \frac{\Delta l}{2m}\right) \quad (\Delta l \text{ above surface}) \tag{2.4}$$

Of course (2.4) is only the Newtonian result and we need GR to get a correct estimate for lowering the box to near a BH, to be done in the next section. That will later lead to an estimate for the BH temperature.

### 3. Energy extraction with a BH

We next calculate the amount of work we can extract from the gravitational field of a BH by lowering the box of mass $\mu$ in Fig. 2 to near the BH surface. This is done much as in section 1. The metric and the geodesic equation will give a potential energy function $V$ much as in classical theory.

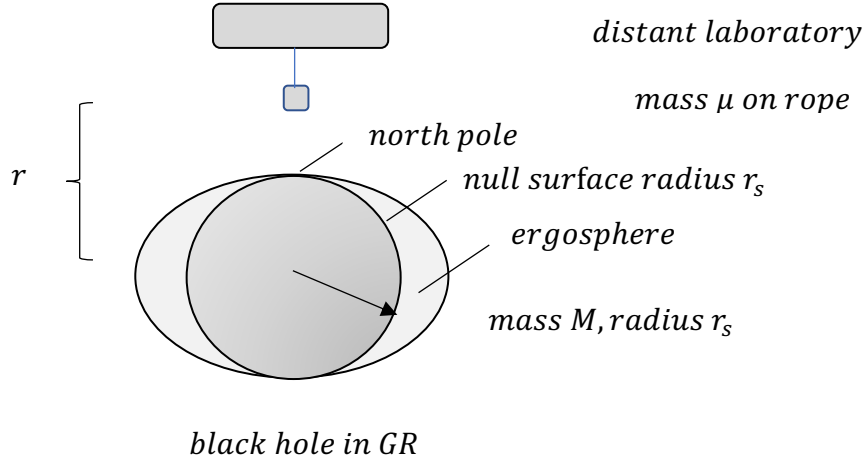

black hole in GR

Figure 2. Thought experiment to extract energy from gravity, using GR. For a non-spinning BH $r_s = 2m$, while for a spinning BH $r_s = m + \sqrt{m^2 - a^2}$.

The analysis may be done by lowering the mass along the north polar axis $\theta = 0$ or in the equatorial plane $\theta = \pi/2$. The equatorial plane was used in Ref. 11. However the polar axis choice is mathematically simpler and we will use it here, as shown in Fig. 2; the two choices give the same result for the temperature.

The Kerr metric for a spinning BH in standard Boyer Lindquist coordinates is [6,14],

$$ds^2 = \left(1 - \frac{2mr}{r^2 + a^2\cos^2\theta}\right)c^2 dt^2 - \left(\frac{r^2 + a^2\cos^2\theta}{r^2 + a^2 - 2mr}\right)dr^2 - (r^2 + a^2\cos^2\theta)d\theta^2$$

$$- \left[(r^2 + a^2)\sin^2\theta + \frac{2mra^2\sin^4\theta}{r^2+a^2\cos^2\theta}\right]d\varphi^2 - 2\left(\frac{2mra\sin^2\theta}{r^2+a^2\cos^2\theta}\right)c\, dt\, d\varphi. \quad (3.1)$$

Here $a$ is a constant parameter with the dimension of a distance (like $m$ and $r$) and is related to the angular momentum by $ma = GJ/c^3$. A dimensionless parameter a* is also widely used in the literature, defined by $a^* = a/m$.

Along the polar axis $\sin\theta = 0$ so the metric is remarkably simple and quite similar to the Schwarzschild metric

$$ds^2 = \left(1 - \frac{2mr}{r^2+a^2}\right)c^2 dt^2 - \left(1 - \frac{2mr}{r^2+a^2}\right)^{-1} dr^2.$$

$$= \left(\frac{r^2+a^2-2mr}{r^2+a^2}\right)c^2 dt^2 - \left(\frac{r^2+a^2-2mr}{r^2+a^2}\right)^{-1} dr^2 \quad (3.2)$$

Moreover it is obvious that geodesic motion occurs along the polar axis.

The metric (3.2) leads to a geodesic equation for the radial velocity, $\dot{r} = \frac{dr}{ds}$, and that equation leads to a definition of the potential $V$ [6,14]. For the case of polar motion (and thus zero angular momentum) these are

$$\dot{r}^2 = \frac{E^2}{\mu^2 c^4} - \frac{V^2}{\mu^2 c^4}, \quad V = \sqrt{1 - \frac{2mr}{r^2 + a^2}} \, \mu c^2, \tag{3.2}$$

Here $E$ is the constant total energy parameter of the mass $\mu$, and we have chosen the positive sign for $V$. (See the appendix for a further discussion of the potential.)

Note the weak field limit for the case of zero spin, $a = 0$; the potential $V$ and its weak field limit are,

$$V = \sqrt{1 - \frac{2m}{r}} \, \mu c^2 \cong \mu c^2 \left(1 - \frac{m}{r}\right) = \mu c^2 - \frac{GM\mu}{r}, \tag{3.3}$$

This is merely the rest energy plus the Newtonian potential, as should be expected.

Consider now the box in Fig. 2 being lowered from a large distance to radius $r$. As in (2.2) the energy extracted will be the difference in the potential energy $V$ in (3.2), so the work done is

$$W = V(\infty) - V(r) = \mu c^2 \left(1 - \sqrt{1 - \frac{2mr}{r^2 + a^2}}\right) = \mu c^2 \left(1 - \sqrt{\frac{r^2 + a^2 - 2mr}{r^2 + a^2}}\right) \tag{3.4}$$

The BH surface, that is the outer null surface of the Kerr metric, is given by [6,14]

$$r^2 + a^2 - 2mr = 0, \quad \text{thus } r_s = m + \sqrt{m^2 - a^2}. \tag{3.5}$$

Thus if we lower the box to the BH surface we obtain work equal to the total rest energy,

$$W = \mu c^2 \quad \text{(rest energy at surface)}. \tag{3.6}$$

This is twice the result we obtained for the classical Newtonian case in (2.3).

Note that the potential (3.2) is derived by considering a body in geodesic motion, and we have assumed that it applies also to an object being lowered very slowly. It is not difficult to show that this is true, which is done in the appendix.

For the energy analysis to follow below we need to calculate $W$ if the box is lowered to *near* the BH, to a small coordinate distance $\Delta r = r - r_s$. Substitution of this into (3.4) and (3.5) gives the approximations, in terms of the important parameter $K$,

$$r^2 + a^2 - 2mr \cong 2 \Delta r \sqrt{m^2 - a^2}$$

$$\frac{r^2 + a^2 - 2mr}{r^2 + a^2} \cong \frac{2 \Delta r \sqrt{m^2 - a^2}}{2mr_s} = \frac{\Delta r}{2m} \frac{2\sqrt{m^2 - a^2}}{(m + \sqrt{m^2 - a^2})} = \frac{\Delta r}{2m} K, \quad K \equiv \frac{2\sqrt{1 - a^2/m^2}}{1 + \sqrt{1 - a^2/m^2}} \tag{3.7}$$

Thus the work $W$ is, from (3.4),

$$W = \mu c^2 \left(1 - \sqrt{\frac{\Delta r}{2m}}\sqrt{K}\right), \qquad \text{(near to BH surface)} \qquad (3.8)$$

This is expressed in terms of the coordinate distance $\Delta r$ above the BH, but for our energy analysis in section 4 we will need to express it in terms of the corresponding *physical* distance $\Delta l$; this is easily done using GR. The metric allows us to relate a physical radial distance differential $dl$ to the coordinate differential $dr$, using the radial metric component, according to

$$dl = \sqrt{|g_{11}|}dr \quad \text{with} \quad \sqrt{|g_{11}|} \cong \frac{1}{\sqrt{K}}\sqrt{\frac{2m}{\Delta r}}, \qquad (3.9)$$

which follows from (3.2) and (3.7). From this the small physical distance $\Delta l$ follows by a simple integration,

$$\Delta l = \int_0^{\Delta r} \sqrt{|g_{11}|}dr = \frac{2\sqrt{2m}\sqrt{\Delta r}}{\sqrt{K}} \qquad (3.10)$$

Combining (3.8) and (3.10) we obtain the desired work $W$ in term of $\Delta l$ in remarkably simple form,

$$W = \mu c^2 \left(1 - \frac{\Delta l}{4m}K\right), \quad K = \frac{2\sqrt{1-a^2/m^2}}{1+\sqrt{1-a^2/m^2}}. \qquad (3.11)$$

Here we have repeated the definition of the important parameter $K$. This is the key relation of this section. It is the basis for the energy analysis in section 4 that leads to the BH temperature. Note that we may think of the quantity in parentheses as the efficiency of the process.

It is important to note that the energy extraction process we have discussed here is *not* related to the Penrose process, by which energy may be extracted from the ergosphere by the interaction of particles.[15]

### 4. View as a Carnot heat engine, the BH temperature

If we combine the above result (3.11) with the Carnot theorem for the efficiency of a heat engine we can estimate the temperature of a BH in a way that uses the minimum of theory - only GR and thermodynamics. The only place $\hbar$ enters is via the Planck expression for the energy of black body radiation.

We now view the system in Fig. 2 as a Carnot heat engine, as in Fig. 3; it uses a hot reservoir at temperature $T_h$ far from the BH, and the BH itself as a cold reservoir.[6.6] At the large distance from the BH we fill the movable box in Fig. 2 with thermal radiation from the hot reservoir and close it; we take the box to be rigid and insulated and sealed so no radiation can escape. Call the full box mass $\mu$ and assume its empty mass is negligible. Then, as in section 3, we lower the box very slowly, with the rope, to do work on some apparatus in the laboratory. We stop the box near the BH surface at a distance $\Delta l$, and open it to let the radiation fall out and be absorbed by the BH. Then we raise the empty box to the starting position and continue. The work done in the process is that given in (3.11).

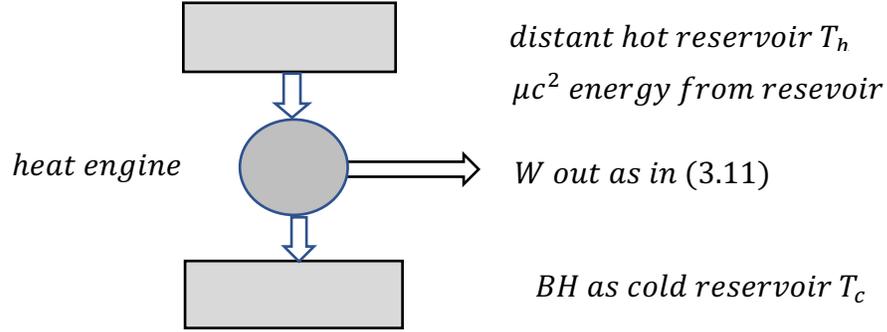

Figure 3. Carnot engine which extracts work $W$. See thermodynamics texts such as Refs. 10 -11.

Recall that the ideal Carnot efficiency $e$, which is the work $W$ divided by the total energy given up by the hot reservoir, is given by [9]

$$e = \frac{W}{\mu c^2} = 1 - \frac{T_c}{T_h}. \qquad \text{(ideal Carnot efficiency)} \qquad (4.1)$$

There is no apparent energy dissipation in our system, so we may equate the efficiency (3.11) with the ideal Carnot efficiency in (4.1) and find

$$\frac{T_c}{T_h} = \frac{\Delta l}{4m} K . \qquad \text{(near BH surface)} \qquad (4.2)$$

If the box could be lowered to the BH surface at $\Delta l = 0$ the ideal efficiency would be unity, corresponding to $T_c = 0$.

But the box cannot be lowered to the surface if it is to contain the thermal radiation from the hot reservoir; we need to estimate how close to the BH we can lower it. This will give an effective BH temperature $T_c$ with the use of (4.1). If we are to load the box with thermal radiation at $T_h$ it must be large enough to contain the relevant photon wavelengths. According to the Planck relation for photons in a black body the average photon energy is of order $h\nu = T_h$ which determines a "characteristic wavelength"

$$h\nu = T_h \quad \text{so} \quad \lambda \cong \alpha \frac{\hbar c}{T_h}. \qquad \text{(temperature in units of energy)} \qquad (4.3)$$

Here $\alpha$ is a dimensionless parameter of order 1. Thus the box must be at least of order

$$\Delta l \cong \lambda \cong \alpha \frac{\hbar c}{T_h}. \qquad \text{(physical box size)} \qquad (4.4)$$

As we noted previously we take the box to be made of very rigid material so its physical size remains constant as it is lowered to near the BH. As we also noted previously, we make the box insulating so its mass and the wavelengths of the black body radiation in (4.4) will not change as it is lowered.

From (4.2) and (4.4) we finally obtain the BH temperature

$$\frac{T_c}{T_h} = \frac{\Delta l}{4m} K = \alpha \frac{\hbar c}{4m\, T_h} \quad , \quad T_c = \alpha \frac{\hbar c}{4m} K = \alpha \frac{\hbar c^3}{4GM} K \,. \tag{4.5}$$

Hawking obtained his temperature for a non-spinning BH, that is $K = 1$,

$$T_H = \frac{\hbar c^3}{8\pi GM}. \quad \text{(Hawking temperature)} \tag{4.6}$$

Thus our estimate (4.5) agrees with the Hawking result for the parameter choice $\alpha = 1/2\pi$.

Our final result is (4.5), which we now rewrite as

$$T_c = \frac{\hbar c}{8\pi\, m} K = \frac{\hbar c^3}{8\pi GM}\left[\frac{2\sqrt{1-a^2/m^2}}{1+\sqrt{1-a^2/m^2}}\right] = T_H\left[\frac{2\sqrt{1-a^{*2}}}{1+\sqrt{1-a^{*2}}}\right], \qquad a^* = a/m. \tag{4.7}$$

This agrees with the standard result obtained using GR and quantum field theory. [3]

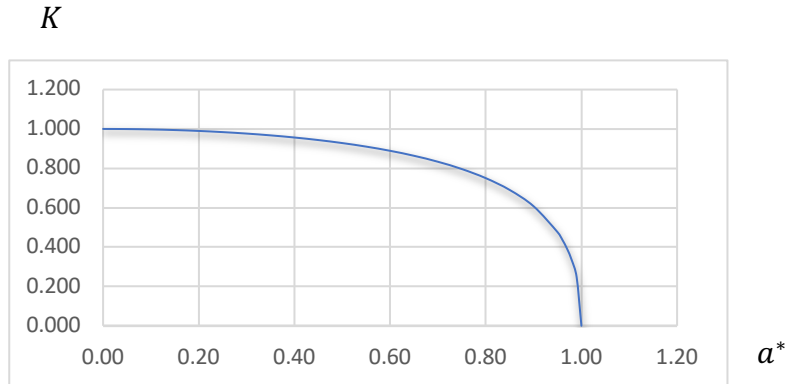

Figure 4. Temperature of the spinning BH (in units of $T_H$) as a function of $a^*$.

The dependence of the temperature on the spin parameter $a^*$, described by $K$, is shown in Fig. 4. Notice that the effect of spin is not large until the angular momentum parameter is near the allowed maximum of $a^* = 1$.[6,11] The temperature drops to $0.5\, T_H$ only at $a^* \cong 0.95$.

A BH with maximum spin $a^* = 1$ and zero temperature should not radiate at all and would be stable against evaporation. It might thus be an interesting candidate for the cosmological dark matter, analogous to the primordial BH remnants discussed in Ref. 12, which we will discuss further in section 5.

Some specific numbers will give a quantitative idea of the relevance of the BH temperature to astrophysics.[6] A solar mass BH would have a Hawking temperature of $6.2 \times 10^{-8}$ K. This is vastly less than the CMB temperature of 2.7 K of the current universe, so such a BH would emit negligible energy compared to what it absorbs. On the other hand, when the CMB was emitted at the recombination time of about 380,000 yr. the ambient temperature was about 3,000 K; a BH of mass much less than about $10^{19}$ $kg$ would have had a higher temperature and would have radiated and evaporated.

**5. Relevance to cosmology**

The prediction that BHs have a temperature inversely proportional to their mass is quite remarkable. The expected behavior of a low mass BH is roughly as follows: if its temperature is less than the ambient temperature it should radiate photons like any black body, and its mass should thus decrease.[6,12] As its mass decreases its temperature should increase, leading to explosive feedback and evaporation according to (4.7). At sufficiently low mass and high temperature it should also radiate massive particles whose rest mass energy is less than about $kT$; indeed it should radiate every type of particle in nature. As it approaches the Planck mass its fate is quite uncertain since diverse quantum effects should become important, and it may either vanish entirely or leave a stable remnant due to quantum effects. Ref. 12 details one scenario leading to a Planck scale remnant. In either case there should be a final intense burst of energy, which would give an almost unique window on physics at the Planck scale. Alas, no such event has yet been observed.[13,16]

Note that according to the BH temperature (4.7) the above comments apply only to a BH which has a spin parameter that is not too close to the maximum $a^* = 1$. BHs with the maximum spin would be stable against radiative decay.

The possibility that the above scenario may result in small BH remnants is of further interest. In the standard cosmological model the universe is made up of about 70% dark energy, 25% dark matter, and only 5% ordinary visible matter.[6] The physical nature of dark matter is not yet known and its existence is evidenced entirely by its gravitational effects.[17,18] Thus it is the focus of much work in both observation and theory. See Ref. 19 for a discussion of possible candidates for the dark matter. In particular the small remnant BHs, perhaps with Planck scale masses, are one of the interesting possible candidates[12,19]. Such dark matter particles would have only weak gravitational interactions, and due to their comparatively large masses they would have very low number density, thus making them extremely difficult to detect experimentally, that is likely to be much more difficult than any other candidates.

**6. Conclusions**

We have shown that the temperature of a spinning Kerr BH can be calculated approximately using GR and black body thermodynamics. The dependence of the temperature on the spin angular momentum that we obtain agrees with the standard result but makes it more accessible to students and nonexperts in quantum field theory. Thus the widespread expectation by theorists that BHs should have a nonzero temperature may be further justified despite the absence of observational evidence.

The fact that low mass BHs should evaporate explosively makes them of interest in fundamental physics at the Planck scale. Finally, BH remnants are a possible candidate for the cosmological dark matter.

**Appendix**

Here we will obtain the potential used in section 3 from a different point of view. Our fundamental physics input is that the equation of motion for a body undergoing motion under both gravity and an external *non-gravitational* force is a geodesic equation with an added term $\mathcal{F}^\gamma$,

$$\frac{D}{Ds}\dot{x}^\gamma = \ddot{x}^\gamma + \Gamma^\gamma_{\alpha\beta}\dot{x}^\alpha\dot{x}^\beta = \mathcal{F}^\gamma \qquad (\mathcal{F}^\gamma \text{ effect of rope holding up the body}). \qquad (a.1)$$

This has the same form as that which describes the motion of a body acted on by gravity and electromagnetism.[20] For the case of the body in Fig. 1 with mass $\mu$ being lowered very slowly we have $x^1 = r$ and $\dot{x}^1 = \ddot{x}^1 = 0$ so (a.1) becomes

$$\Gamma^1_{00}\dot{x}^0\dot{x}^0 = \mathcal{F}^1. \qquad (a.2)$$

This is the static force holding the body nearly at rest. For the polar axis BH metric in (3.2) the quantities in (a.2) are

$$\Gamma^1_{00} = \frac{1}{2}g^{11}(g_{01,0} + g_{10,0} - g_{00,1}) = -\frac{1}{2}g^{11}g_{00,1}, \qquad (a.3)$$

$$(\dot{x}^0)^2 = \frac{1}{g_{00}}, \quad g^{11} = -g_{00}, \quad g_{00,1} = m\frac{r^2-a^2}{(r^2+a^2)^2} \qquad (a.4)$$

Thus we have

$$\mathcal{F}^1 = g_{00,1} = m\frac{r^2-a^2}{(r^2+a^2)^2} \qquad (a.5)$$

It is worth noting that for the case of zero spin, $a = 0$, this gives

$$\mathcal{F}^1 = \frac{m}{r^2} = \frac{GM}{c^2 r^2} = \left(\frac{GM\mu}{r^2}\right)\frac{1}{\mu c^2} \qquad \text{(zero spin case)} \qquad (a.6)$$

It is clear that $\mathcal{F}^1$ is a force $F$ per unit rest energy,

$$F = \mu c^2 \mathcal{F}^1 = \mu c^2 m\frac{r^2-a^2}{(r^2+a^2)^2} = GM\mu\frac{r^2-a^2}{(r^2+a^2)^2} \qquad (a.7)$$

With this expression for the force $F$ we may calculate the work $W$ done by slowly lowering the body as the integral of the force $F$ along the distance to the BH. However the distance differential should be the physical $dl$ rather than the coordinate $dr$; the two are related in (3.9). Thus we obtain

$$W(r) = \int_\infty^r F\, dl = -\int_r^\infty F\frac{dl}{dr}dr = -\mu c^2 m\int_r^\infty \frac{r^2-a^2}{(r^2+a^2)^2\sqrt{1-\frac{2mr}{r^2+a^2}}}dr$$

$$= \mu c^2\left(1 - \sqrt{1 - \frac{2mr}{r^2+a^2}}\right). \qquad (a.8)$$

This is the same as (3.4) which we obtained in section 3. Thus the potential $V$ for a slowly lowered body and that for free fall are the same.

## References

[1] S. W. Hawking, "Particle creation by black holes," Comm. Math. Phys. **43**, 199 (1975).

[2] N. D. Birrel and P. C. W. Davies, *Quantum Fields in Curved Space*, chap 8 (Cambridge University Press, 1982).

[3] R. Blandford and K. Thorne, *Modern Classical Physics*, chap 21 (Princeton University Press, 2017). The temperature is here related to the acceleration of gravity at the BH surface.

[4] Y. C. Chou, "A radiating Kerr black hole and Hawking radiation," Heliyon 6 (2020).

[5] W- X Chen, J-X Li, J-Y Zhang, "Calculating the Hawking temperature … ," Int. J. Th. Phys. **62** 5042, (2023).

[6] R. J. Adler, *General Relativity and Cosmology, a First Encounter*, chap 19 (Springer, 2021).

[7] M. K. Parikh and F. Wilczek, "Hawking radiation as tunneling," Phys. Rev. Lett. **85**, 5042 (2000);

[8] T. S. Bunch, *J. Phys. A: Math Gen.* **14**, L139 (1981).

[9] D. V. Schroeder, *An introduction to Thermal Physics,* chap 4 (Addison Wesley, 2000).

[10] H. Ohanian and R. Ruffini, *Gravitation and Spacetime*, 2nd ed., chap 8 (W. W. Norton, 1994).

[11] R. J. Adler, "On the temperature of a rotating black hole," Int. J. Mod. Phys D, **34**, 16 (2025).

[12] R. J. Adler and P. Chen and D. I. Santiago, "The generalized uncertainty principle and black hole remnants," Gen. Rel. Grav. 33 (2001).

[13] C. Yang et. al. , "Search for the Hawking radiation of primordial black holes," JCAP 10 (2024). No confirmed observation of Hawking radiation has been reported.

[14] C. W. Misner and K. S. Thorne and J. A. Wheeler, *Gravitation*, chap 33 (W. H. Freeman, 1973).

[15] R. Penrose and R. M. Floyd, "Extraction of rotational energy from a black hole," Nature Phys. **229**, 177 (1971).

[16] Zhen Cao et. al. , "All sky search for individual black hole bursts with LHAASO," Phys. Rev. Let. 135, (2025).

[17] Strong evidence for dark matter has existed since the 1930s, notably F. Zwicky, " Die Rotverschiebung von extragalaktischen Nebeln,": Helv. Phys. cta 6 (1933).

[18] V. Rubin and W. K. Ford, "Emission-line intensities and radial velocities … " Astr. Phys. J. 159 (1970). There is now a vast literature on dark matter properties and the nature and the search for it in experiments.

[19] N. Bozorgnia, "Dark matter candidates and searches," arxiv hep-ph, 2410.23454 (2025). The Wikipedia article on Dark Matter also includes an extensive list of references.

[20] R. J. Adler and M. Bazin and M. M. Schiffer, *Introduction to General Relativity; 2nd. edition*, chap 7 (McGraw Hill, 1975).